\documentclass[aps,prl,twocolumn,showpacs]{revtex4}

\usepackage{graphicx}

\begin{document}

\newlength\FigWidth \FigWidth 90 true mm

\title{Nanoscale Zeeman localization of charge carriers in diluted
  magnetic semiconductor-permalloy hybrids}

\author{Mona Berciu$^1$ and Boldizs\'{a}r Jank\'{o}$^{2,3}$}

\affiliation{$^1$Department of Physics and Astronomy, University of
  British Columbia, Vancouver BC, V6T 1Z1, Canada}

\affiliation{$^2$Materials Science Division, Argonne National
  Laboratory, Argonne, Illinois 60439}

 \affiliation{$^3$Department of
  Physics, University of Notre Dame, Notre Dame, Indiana 46556}

\begin{abstract}
We investigate the possibility of charge carrier localization in
magnetic semiconductors due to the presence of a highly
inhomogeneous external magnetic field. As an example, we study in
detail the properties of a magnetic semiconductor-permalloy disk
hybrid system. We find that the giant Zeeman response of the
magnetic semiconductor in conjunction with the highly non-uniform
magnetic field created by the vortex state of a permalloy disk can
lead to {\em Zeeman localized} states at the interface of the two
materials. These trapped states are chiral, with chirality
controlled by the orientation of the core magnetization of the
permalloy disk. We calculate the energy spectrum and the
eigenstates of these Zeeman localized states, and discuss their
experimental signatures in spectroscopic probes.
\end{abstract}
\maketitle


Diluted magnetic semiconductors (DMS) based on III-V alloys doped
with Mn have attracted a lot of interest recently, due to their
relatively high Curie temperatures T$_{\rm c}$ (110 K for
Ga$_{0.95}$Mn$_{0.05}$As)\cite{Ohno}, below which they exhibit
ferromagnetic  order. In the ferromagnetic state, the charge
carriers are spin-polarized, making these materials ideal sources
of spin-polarized currents. To date, most suggested applications
involving DMS are based on this property and therefore are
restricted to operation in a range of temperatures T $\ll$ T$_{\rm
c}$.  On the other hand, in the paramagnetic phase T $>$ T$_{\rm
c}$ both the III$_{1-x}$Mn$_x$V and the more established
II$_{1-x}$Mn$_x$VI DMSs (which  have even  lower critical
temperatures) show giant Zeeman response to external magnetic
fields.  In our opinion, this can lead to interesting applications
in the paramagnetic state, and consequently at rather elevated
temperatures.

Convincing experimental evidence for Zeeman splitting in the range of
$30 \ {\rm meV}$ for external fields of a few Tesla is provided by
photoluminesce spectroscopy studies \cite{margaret}. Even relatively
small external magnetic fields of $0.1-0.5$ T can easily lead to a 15
meV splitting of electronic energy levels \cite{Jacek}. Comparing this
to the vacuum Zeeman splitting of $\sim$ 0.06 meV (for $B=0.5T$)
suggests that the effective gyromagnetic ratio of charge carriers in
diluted magnetic semiconductors is $g > 500$. The origin of this
hugely enhanced Zeeman effect is attributed \cite{Jacek} to the strong
magnetic coupling $ \sum_{i}^{}J_{sp-d}(\vec{r}-\vec{R}_i)\vec{s}\cdot
\vec{S}_i$ between the spin $\vec{s}$ of the charge carrier and the
spins $\vec{S}_i$ of the Mn located at $\vec{R}_i$. In the
paramagnetic state, a small magnetic field $\vec{B}$ induces a
magnetization $\langle\vec{S}_i\rangle \sim \chi \vec{B} $ of the Mn
spins, resulting in an effective Zeeman-like $\vec{s}\cdot \vec{B}$
coupling between the charge carrier spin and the magnetic field, in
addition to the regular Zeeman coupling $-g_0\mu_B \vec{s}\cdot
\vec{B}$ present in non-magnetic semiconductors.  The scale
of this additional coupling is set by the large exchange energy
$J_{sp-d}$ and results in a large effective $g$-value.  This also
implies that $g(T)$ has a strong T-dependence through the magnetic
susceptibility, and therefore can be tuned over a large range of
values. This scenario is strongly supported by magneto-optical
absorption measurements \cite{margaret2} of the Zeeman splitting at
the band edge, which clearly exhibits a Brillouin-type dependence on
the magnetic field.

The presence of giant Zeeman response in DMS  implies that a moderate
external magnetic field 
with a strong spatial variation on nanometer scale can be a very
effective confining agent for spin-polarized charge carriers in
these systems. Provided that such highly inhomogeneous external
fields can be created and controlled adequately in a DMS,
Zeeman-induced localization presents a new route for {\em
manipulation of spin-polarized charge carriers at relatively high
temperatures}.

Non-uniform magnetic fields with nanoscale spatial variations are
known to appear in a variety of systems, such as the Abrikosov flux
lattice \cite{algas} and arrays of nanoscale holes in superconducting
films \cite{Vitali}. However, one of the most promising possibilities,
which we investigate in this Letter, is provided by  the magnetic vortex
state \cite{cowburn} of nanoscale magnetic disks of ferromagnetic
permalloy Ni$_{80}$Fe$_{20}$, Co or Fe. In the remanent state of such
nanomagnets, the local magnetization near the perimeter of fairly
thin disks has an in-plane vortex-like arrangement. This is
energetically more favorable than a single ferromagnetic domain, since
the exchange energy lost due to the gradual vortex rotation is more
than compensated by the cancellation of the total dipole energy.  This
pattern is maintained for almost the entire volume of the
disk. However, near the center of the disk, exchange
interaction wins over dipole-dipole interaction and shape anisotropy,
and the local magnetization is forced out of the plane of the disk.
What is remarkable about this topological singularity in the
magnetization is the extremely short length scales and high fields
involved: Very recent experimental investigations indicate that the
radius of the magnetic core is about 30nm in permalloy disks
\cite{Shinjo,Raabe} and 10 nm in Fe disks \cite{Wachowiak}, with
maximum field values at the core  in the 0.5-1.0 Tesla
range. In this paper we show that such highly
inhomogeneous magnetic fields provides a very effective localization
agent for charge 
carriers in a system with large Zeeman effect.

\begin{figure}
\centering \includegraphics[angle=270,width=\FigWidth]{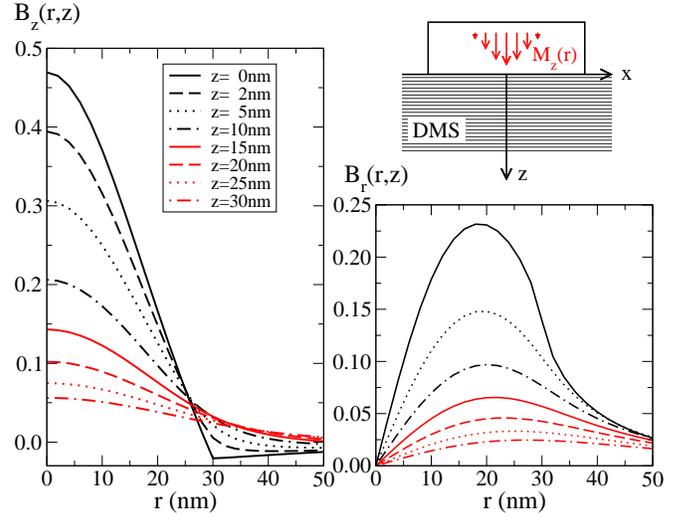}
\caption{\label{fig1} $B_z(r,z)$ (left) and $B_r(r,z)$ (right) in
units of $\mu_0M_0\approx 1$ T, plotted as a function of $r$ for
several values of $z$, for a disk of height $d=50$ nm and a core
radius $\rho_c$ = 30 nm. The discontinuity in the slope of $B_z(r,0)$
at $r=\rho_c$ is a consequence of the discontinuity of $dM_z(r)/dr$ at
$r=\rho_c$ [see Eq. (\ref{7})]. }
\end{figure}

We investigate the properties of magnetic disks for which the height $d$ is
small compared to radius $R$. In this limit, the disks exhibit
vortex magnetization of the following general type \cite{book}
\begin{equation}
\label{1}
\vec{M}(\vec{r}) = M_{\phi}(r)\vec{e}_\phi + M_z(r)\vec{e}_z
\end{equation}
where cylindrical coordinates $\vec{r} = (r,\phi,z)$ are used.
[Note that $r \ne |\vec{r}|$].  Using Maxwell's equations
\cite{Jackson}, we find the magnetic field created by a
magnetization of the type described in Eq. (\ref{1}). For the
setup shown in Fig. \ref{fig1}, the magnetic field created in the
DMS layer ($z>0$) is $\vec{B}(\vec{r}) = \vec{b}(\vec{r})-
\vec{b}(\vec{r}+d \vec{e}_z)$, where, in cylindrical coordinates
$\vec{r} = (r,\phi,z)$:
$$ b_r(r,z) = {\mu_0\over 2\pi r} \int_{0}^{\rho_c} {{ d\rho \rho
M_z(\rho)} \over \sqrt{(r+\rho)^2+z^2}} \left\{ K\left(f(r,\rho,z)
\right) \right.
$$
\begin{equation}
\label{2}
\left.  - E\left(f(r,\rho,z)\right)\cdot [\rho^2 + z^2
  -r^2]\cdot[(r-\rho)^2 + z^2]^{-1} \right\}
\end{equation}
\begin{equation}
\label{3}
b_\phi(r,z) = 0
\end{equation}
and
\begin{equation}
\label{4}
b_z(r,z) = { \mu_0 z\over \pi} \int_{0}^{\rho_c} { d\rho\rho M_z(\rho)
 \over (r-\rho)^2 + z^2} \cdot
 {E\left(f(r,\rho,z)\right)\over\sqrt{(r+\rho)^2+z^2}}
\end{equation}
while the corresponding magnetic vector $\vec{A}(\vec{r}) =
\vec{a}(\vec{r})- \vec{a}(\vec{r}+d \vec{e}_z)$, in the Coulomb gauge
$\nabla \cdot \vec{A}=0$, is given by
\begin{equation}
\label{5}
a_r(r,z) = a_z(r,z) = 0
\end{equation}
and
$$ a_\phi(r,z) = - {\mu_0\over 2\pi}{z \over r} \int_{0}^{\rho_c}
{d\rho\rho M_z(\rho) \over (r+\rho)\sqrt{(r+\rho)^2+z^2}}\times
$$
\begin{equation}
\label{6}
 \left[ (r-\rho) \Pi\left({4r\rho \over (r+\rho)^2}, f(r,\rho,z)
\right) + (r+\rho) K\left(f(r,\rho,z) \right) \right]
\end{equation}
Here, $f(r,\rho,z) = \sqrt{(4r\rho)/ [(r+\rho)^2+z^2]}$ and the
elliptic functions $K(k),E(k)$ and $\Pi(\nu,k)$ are defined in
Ref. \onlinecite{ell}. The fact that only $M_z(r)$ enters these
equations is expected, since the field lines induced by
$M_\phi(r)$ are closed, and therefore this component does not
contribute to the magnetic field outside the magnetic disk. Recent
micromagnetic simulations have shown \cite{cit1,Wachowiak} that the
magnetization inside the disk is well fitted by the simple
parameterization \cite{book,Aha,Usov}:
\begin{equation}
\label{7}
|M_z(r)|/M_0 = \left\{
\begin{array}[c]{ll}
  {(\rho_c^2 - r^2 ) / (\rho_c^2 + r^2 )} \hspace{5mm}&,r \le\rho_c \\
0 &, r > \rho_c \\
\end{array}
\right.
\end{equation}
and $|M_\phi(r)| =\sqrt{M_0^2-M_z^2(r)}$.  Such a structure is also
known as a meron configuration in quantum Hall systems
\cite{Gross}. In permalloy disks, the saturation magnetization is
$\mu_0 M_0 = 1.06$~T and the core radius $\rho_c \approx 30$~nm, if
the disk radius $R \gg \rho_c$ \cite{Raabe}. [The core radius $\rho_c$
decreases with decreasing $d$]. The core magnetization $M_z$ has been
observed experimentally in permalloy disks of typical height $d = 50$
nm and radii $R=0.1- 1~\mu$m \cite{Shinjo} as well as Fe disks with
$d=9$ nm and $R=200-500$~nm \cite{Wachowiak}.

Using Eqs. (\ref{2}-\ref{4}) and (\ref{7}), we find the magnetic
field created by the vortex inside the diluted magnetic semiconductor
to have a typical profile as shown in
Fig. (\ref{fig1}). Inside the DMS layer, $B_z(r,z)$ is largest at
$r=z=0$, and decreases with increasing $r$ and $z$. $B_r(r=0,z)=0$
as expected for a dipole-like field. However, $B_r(r,z)$ reaches a
maximum value which is roughly half of the maximum $B_z(r,z)$
value for the same $z$, at a distance $r$ of the order of
$\rho_c$.  The giant Zeeman effect present in the DMS systems
leads to the possibility of trapping carrier states in the strong
magnetic field near the disk surface.

The eigenstates of a charge carrier placed in such a magnetic field
are given by the Schr\"{o}dinger equation:
\begin{equation}
\label{8}
\left\{{ 1\over 2m} \left[\vec{p} + q \vec{A}(\vec{r}) \right]^2 - g
  \mu_B { \vec{\sigma}\over 2} \cdot \vec{B}(\vec{r}) \right\}
  \phi(\vec{r}) = E\phi(\vec{r})
\end{equation}
where, as already mentioned, the large $g$-factor effectively
accounts for the magnetic exchange of charge carriers with locally
magnetized Mn spins. The Hamiltonian given in Eq.(\ref{8})
neglects electron-electron interactions. To first order, this is
justified if the carrier concentration is either large enough that
screening is very effective, or for low electron concentrations
where only a small number of electrons might be trapped below each
disk. In the latter case, Eq. (\ref{8}) should accurately describe
the strongest-bound state; for higher energy bound states, one
should also include the screening provided by electrons occupying
inner shells. We omit this complication here. Eq. (\ref{8}) also
assumes that the Mn spins create a smooth polarizing field for the
electron, ignoring possible local fluctuations due to the fact
that Mn spins are distributed randomly in the DMS. This so-called
virtual crystal approximation has already been used extensively to
describe DMS systems.

Given the symmetries of the magnetic and vector fields [Eqs.
(\ref{2}-\ref{6})], Eq. (\ref{8}) has solutions of the following
general form:
\begin{equation}
\label{9}
\phi_{m}(r, \phi,z) = \exp{(im\phi)}\left(
\begin{array}[c]{c}
\phi_{\uparrow}^{(m)}(r,z) \\
\phi_{\downarrow}^{(m)}(r,z)\exp{[i\phi]} \\
\end{array}
\right)
\end{equation}
where the angular momentum $m$ is an integer. The extra phase
$\exp{[i\phi]}$ in the spin-down component has two significant
consequences: (1) $\sigma_z$ is not a good quantum number.
[Besides the $\langle \hat{s}_z \rangle \ne 0$ part, the
expectation value of the spin acquires a radial contribution
$\langle \hat{s}_x + i \hat{s}_y \rangle \sim \exp{(i\phi)}$, for
all $m$. This is clearly due to the existence of a radial part
$B_r(r,z)$ in the magnetic field], and (2) the (usually expected)
degeneracy between states with $\pm m$ is now lifted. The reason
is that the presence of the magnetic field breaks the
time-reversal symmetry responsible for this degeneracy. This has
interesting consequences, as described below.

The general form of the wave function given above allows us to
draw immediate conclusions regarding selection rules governing the
electromagnetic response of the electronic states trapped in the
DMS by the vortex magnetic field. Assume that the system is
exposed to monochromatic radiation, and let $\vec{A}_L(\vec{r})$
and $\vec{B}_L(\vec{r})=\nabla\times\vec{A}_L(\vec{r}) $ be the
vector potential, respectively magnetic field associated with it.
Then the Schr\"{o}dinger equation (\ref{8}) acquires three extra
terms, proportional to $\vec{\sigma}\cdot \vec{B}_L(\vec{r})$,
$(\vec{p}+q\vec{A}) \cdot \vec{A}_L(\vec{r})$ and to
$\vec{A}_L^2(\vec{r})$, in order of decreasing magnitude. Assume
that the beam propagates along the $z$-axis and is circularly
polarized, $\vec{B}_L^{(\pm)}(z) \sim (\vec{e}_x \pm i
\vec{e}_y)\exp{(ikz)}$. Using Eq. (\ref{9}) it is straightforward
to find the selection rules $\langle m' | \vec{\sigma}\cdot
\vec{B}_L^{(\pm)}|m\rangle \sim \delta_{m',m\pm 1}$. (The term
$(\vec{p}+q\vec{A}) \cdot \vec{A}_L(\vec{r})$ obeys the same
selection rules, while the $\vec{A}_L^2(\vec{r})$ term induces
two-photon processes but with a vanishingly small probability
\cite{note}).  This selection rule implies that the absorption of
a right (left) circularly polarized photon excites the electron to
a level with an $m$ increased (decreased) by one unit. If levels
with $\pm m$ are no longer degenerate, the system will interact
differently with photons of different circular polarizations.

Let us now use a simplified model to demonstrate the lifting of the
$\pm m$ degeneracy.  First, the terms involving the vector field
$\vec{A}(\vec{r})$ are removed from Eq. (\ref{8}). This is justified
since they are vanishingly small compared to the Zeeman term, due to
the supplementary enhancement of the later by the interaction with the
Mn spins. Second, since we are interested in the most strongly-bound
states, which are likely to be localized at small $(r,z)$ values, we
use Taylor series for the magnetic fields in this region, given (up to
quadratic terms) by:
\begin{eqnarray*}
& &B_r(r,z) = \mu_0M_0 B\left(z\over\rho_c\right){r \over \rho_c} \\ &
&B_z(r,z) =\mu_0M_0\left[A\left(z\over\rho_c\right)- b_3 {r^2\over
\rho_c^2 }\right]
\end{eqnarray*}
where the functions
$$ A(z) = b_1-b_2z +2b_3z^2 \hspace{5mm} \mbox{and}
\hspace{5mm}B(z) ={b_2 \over 2} - 2b_3z $$ have been introduced
for later convenience.  These  fields continue to satisfy the
condition $\nabla\cdot \vec{B}=0$, as necessary. The coefficients
$b_1, b_2$ and $b_3$ are complicated functions of $d/\rho_c$. In the limit
$d/\rho_c\rightarrow\infty$, we find $b_1 = 1/2$, $b_2=(\pi+2)/4$
and $b_3=1$. These asymptotic values turn out to give 
reasonable approximations for the typical experimental parameters  $d =
50~$nm and $\rho_c=10-30~$nm, with relative errors  less than 10\% for
$d/\rho_c = 5/3$ and decreasing very fast to zero for larger
ratios \cite{new}.  As a result, the asymptotic values will  be  used
in the remainder of this  paper.

With these simplifications, we look for eigenfunctions [see Eq.
(\ref{9})] of the following form:
\begin{eqnarray*}
\phi_{\uparrow}^{(m)}(r,z) = a_1\left({ z / \rho_c}\right)
  \left({r\over \rho_c}\right)^{|m|} \exp{\left( - {r^2 \over
  b^2\rho_c^2}\right)} & &\\ \phi_{\downarrow}^{(m)}(r,z) = a_2\left({
  z/ \rho_c}\right) \left({r\over \rho_c}\right)^{|m+1|} \exp{\left( -
  {r^2 \over b^2\rho_c^2}\right)}& &
\end{eqnarray*}
Both components are regular for $r \rightarrow 0$. The functions
$a_1(z/\rho_c)$ and $a_2(z/\rho_c)$ are determined, in dimensionless units
$ z/\rho_c \rightarrow z$, by the set of coupled equations:
\begin{equation}
\label{e1}
\left[{4(|m|\!+\!1) \over b^2}\! -\! {d^2 \over dz^2 }\! -\! \alpha
  A(z) \right]\!\!  a_1 + {s\!-\!1 \over 2}\alpha B(z) a_2 = e a_1
\end{equation}
\begin{equation}
\label{e2}
\left( { 4 \over b^4} - \alpha\right) a_1(z) + {s+1\over 2} \alpha
 B(z) a_2(z) = 0
\end{equation}
\begin{equation}
\label{e3}
\left[{4(|m|\!+\!s\!+\!1) \over b^2}\!-\! {d^2 \over dz^2}\! +\!
  \alpha A(z)\right]\!a_2\! -\!  {s\!+\!1\over 2}\alpha B(z) a_1 = e
  a_2 
\end{equation}
\begin{equation}
\label{e4}
-\left( { 4 \over b^4} + \alpha\right) a_2(z) +{s-1 \over 2}\alpha
 B(z) a_1(z) = 0
\end{equation}

Here, $s=1$ if $m\ge 0$ and $s=-1$ if $m<0$. We define the energy
unit $E_0 = \hbar^2/(2m\rho_c^2)$ and use $e = E/E_0$, while
$2\alpha = g\mu_B \mu_0 M_0/E_0$ is the ratio between the maximum
Zeeman energy and this energy unit. This number is rather large.
If we use $\rho_c = 30$ nm and $\mu_0 M_0 = 1.06$ T (typical values 
for permalloy disks), $m = 0.5 m_0$ (band value for heavy
hole mass in GaAs) and use the estimated effective gyromagnetic
factor to be $g=500$, we find $|\alpha| \approx 175 $.

If $s=+1$, we see from Eq. (\ref{e4}) that there is a non-trivial
solution if and only if $4/ b^4 = - \alpha$, which is possible if
$\alpha < 0$. This is the case when $M_0 < 0$, i.e. the $z$-axis
disk magnetization $M_z$ points away from the DMS surface
\cite{note2}. With $b^2 = \sqrt{4/|\alpha|}$, Eq. (\ref{e2}) has
the solution $a_1(z) = { 1 \over 2} B(z) a_2(z)$.  Using this in
Eq. (\ref{e3}) we find $a_2(z) = \exp{(-kz)}$, where $k > 0$ is
linked to the eigenenergy through $e_m = 2\sqrt{|\alpha|}(m+2)+
\alpha { (6-\pi)(10+\pi) \over 128}- k^2$. Finally, Eq. (\ref{e1})
must be satisfied up to powers of $z^0$ (the magnetic fields, and
therefore the functions $a_1$ and $a_2$, are accurate only up to
$z^2$, meaning that the second derivative of $a_1$ is accurate only up
to $z^0$). This gives $ k= -0.321\sqrt{|\alpha|}+0.127|\alpha|$ (with
$k>0$ satisfied for any $|\alpha| > 1.56$). In other words, up to
a large negative constant the spectrum $e_m = 2\sqrt{|\alpha|}m$
is similar to that of a harmonic oscillator ($s=+1 \rightarrow
m\ge 0$) \cite{note3}. For typical values for the vortex core radius
and effective mass 
(constants which set the energy scale $E_0$ as given above), the
level spacing is in the range of $3-6 \  {\rm meV}$, which should
be accessible by most spectroscopic tools.

To summarize, for $M_z <0$ we find only solutions with $s=+1$ or $m\ge
0$, proving that the $\pm m$ degeneracy is indeed lifted in this
simplified case.  For $M_z>0$, the reverse is true; we find only
solutions with $m<0$. However, these simple solutions only hold for
small values of $m$, where the wave-functions are localized at small
$r$ and $z$ and the Taylor series for the magnetic fields are
valid. To find the true spectrum, one must integrate Eq. (\ref{9})
numerically for the full expression of the magnetic fields; this work
is in progress and the results will be reported elsewhere
\cite{new}. We expect that the degeneracy between $\pm m$ eigenstates
is lifted in the general case as well. As already discussed, this
means that the system will interact differently with photons of
different circular polarization, depending also on the orientation
(sign of) the magnetic disk core magnetization $M_z$.  Such a property
may be useful in designing spintronic devices : the memory bit ($M_z$
up or down) controls the properties of the trapped electronic states,
and therefore the behavior of the whole device. Other geometries, such
as lines of ordered disks, could be used to trap and transport electric
currents with similar non-trivial spin properties.

In conclusion, we suggest a different route to creating spintronic
devices that can operate {\em at room temperature}, by combining
the giant Zeeman effect of diluted magnetic semiconductors in the
paramagnetic state with the highly inhomogeneous magnetic field
created by nanoscale permalloy disks, or other nano-patterned
magnetic material structures. In particular, we have shown that
electronic states can be trapped near the surface of a magnetic
disk. Their properties can be suitably tailored  using
various  diluted magnetic semiconductors and various
types of magnetic disks, as well as by varying the temperature (since
the effective g-factor has significant temperature dependence). 
Since time-reversal symmetry is broken in the presence of the
magnetic field, the system will interact non-trivially with any
other system which has a definite chirality, such as circularly
polarized light. To our knowledge, this is the first time that
Zeeman-induced localization has been suggested and demonstrated
theoretically, or that electronic states with such unusual
spin-polarization have been derived. Experiments to test these
i are currently in progress \cite{Jacek2}.

{\bf Acknowledgements} We thank Gyorgy Csaba, V. Novosad, M.
Grimsditch, J. K. Furdyna and V. Metlushko for useful discussions.
This research was supported by NSERC of Canada (M.B.) and by NSF
NIRT award DMR 02-10519 and the Alfred P. Sloan Foundation (B.J.).  We
also gratefully acknowledge the hospitality of the Argonne National
Laboratory (M.B. and B.J) and  the Aspen Center for Physics (M.B.),
where parts of this work were carried out.


\begin{thebibliography}{99}
\bibitem{note}
If the beam is shined along the DMS
  surface (propagation in the $xOy$ plane) transitions are allowed
  between any two states $m$ and $m'$, i.e there are no selection
  rules.
\bibitem{note2} $\alpha <0$ is also realized if $M_z >0$, but $g<0$,
i.e. charge carriers are antiferromagnetically coupled to Mn
spins. In the following we use the convention that $g>0$.
\bibitem{note3} Proper boundary conditions must be
  imposed before the exact eigenstates can be found. These depend
  on the nature of the DMS-permalloy interface.
\end{thebibliography}
\end{document}